\documentclass[11pt]{article}
\usepackage[dvips]{graphicx}
\def\PLB{{\em Phys. Lett.}B}

\def\be{\begin{equation}}
\def\ee{\end{equation}}
\def\bea{\begin{eqnarray}}
\def\eea{\end{eqnarray}}

\input epsf.tex

\title{
\rightline{IFT-UAM/CSIC-99-33}
\rightline{hep-th/9909080}
\vspace{1cm}
T-duality and Gauge Theories from Near Horizon Dp-branes}
\author{Pedro Silva\footnote{E-mail:psilva@delta.ft.uam.es}\\
\\
\\
{\it Instituto de Fisica Teorica, C-XVI, Universidad Autonoma de Madrid}\\
{\it E-28049-Madrid, Spain\footnote{Unidad de Investigacion asociada al centrode Fisica Miguel Catalan (C.S.I.C.)}}\\
{\it Physics department, University of Newcastle Upon Tyne}\\
{\it NE1 7RU, UK}\\
{\it Depto de Fisica Teorica, C-XI, Universidad Autonoma de Madrid,}\\
{\it E-28049-Madrid, Spain}}

\begin{document}
\maketitle
\begin{abstract}{We study the significance of T-duality in the context of the gravitational description of gauge theories. We found that T-duality relates the deferents points of the moduli of a given  gauge theory always far from the conformal fixed point. Also the described gauge theories seems to flow naturally to the able conformal points, those that naturally saturate all the possible known examples of near horizon geometries. Supersymmetry properties and T-duality breaking of it are discuss.}
\end{abstract}

\pagebreak
\newpage

%%%%%%%%%%%%%%%%%%%%%%%%%%%%%%%%%%%%%%%%%%%%%%%%%%%%%%%%%%%%%%%%%%%%%%%%%%%%%%%

\section{Introduction}

%%%%%%%%%%%%%%%%%%%%%%%%%%%%%%%%%%%%%%%%%%%%%%%%%%%%%%%%%%%%%%%%%%%%%%%%%%%%%%%

A natural question to ask in the general framework of string descriptions of gauge theories, is the quantum field theoretical meaning of pure string symmetries(for a review closely related see \cite{ces1}). In the celebrated AdS/CFT correspondence the $SL(2,Z)$ S-duality of the Type IIB-string can be interpreted as the string version of the well known Monton Olive $SL(2,Z)$ duality of $N=4$ SYM \cite{sus1}. The quantum field theory meaning of T-duality is however a bit less clear since T-duality interpolates different D-branes and therefore tends to define maps between Yang Mills theories in different dimensions. Moreover T-duality transformation generally produce explicit breaking of supersymmetry.

The simplest possible way to address this questions on T-duality in a purely quantum field theoretical framework is of course by defining the quantum field theory using the near horizon (NH) limit of D-branes metrics and working out in this limit T-duality transformations. Nevertheless the gravitational description of the gauge theories related to the non-conformal D3-brane are not so well understood. The works on AdS/CFT studying holography are not applicable (at least directly) to these cases as the geometries involved are not $AdS$.

To proceed with this studies, we concentrate on the T-duality between the different D-branes, and consider the commutativity between the NH limit and the T-duality transformation. Once this is done, the NH geometries for Dp-branes, with $p \neq 3$ are rewritten in terms of ``$AdS$ variables''. Then what we have found is a metric with an  $AdS$ term plus other factors, that in the allowed limits of the original papers \cite{mal1}, correspond to small compact dimension. Hence, recovering in this way an induce conformal structure. As a direct application, by looking into the induced action of $SO(2,4)$ on the T-dual geometries of $AdS_5$, we have obtained a new kind of symmetry group, baptised as the generalised conformal group (GCG), a sort of generalisation of the conformal transformations found by Jevicki and Yoneya in a completely independent manner in a recent paper \cite{jevi1}. 

In the following we study the T-duality transformation for the NH limit of the D-branes, and find under what specific limits the duality is realised. In particular as an example, the D0-brane case is considered. Then a few conclusions and remarks are stated giving an interpretation for the holography conjecture in these cases. Then the study on the moduli space of the corresponding SYM is given,  resulting in new approaches to the conjecture, interrelating NH D-branes and NH M-branes. Also the appropriated discussion on supersymmetry enhancement is include.

%%%%%%%%%%%%%%%%%%%%%%%%%%%%%%%%%%%%%%%%%%%%%%%%%%%%%%%%%%%%%%%%%%%%%%%%%%%%%%%

\section{T-duality and Near Horizon geometries}

%%%%%%%%%%%%%%%%%%%%%%%%%%%%%%%%%%%%%%%%%%%%%%%%%%%%%%%%%%%%%%%%%%%%%%%%%%%%%%%

To study the holographic nature of the Maldacena conjecture, we must make contact with the D3-brane case, as it is the only case where we believe that we know the mathematical structure underlying the conjecture. We also know that the D-branes are related by T-duality, and that this duality is also realised on the low energy solitonic solutions of supergravity and in the corresponding Born-infeld type action describing the internal degrees of freedom of the D-brane. Therefore an obvious line of approach to understand the mathematical structure of the above conjecture for $p \neq 3$ is to prepare the Dp-branes for T-duality i.e. to set them up wrapped on a torus $T^S$. Then, take the NH limit and check the T-duality of this geometry. The result of this type of analysis should be a precise dictionary from $Ads$ variables in the D3-brane case and the variables in the other NH geometries, allowing a careful inspection on both structures.

Consider the case of $N$ D0-branes on a torus $T^3$. The Supergravity solution is given by the equation,
\begin{eqnarray}
&&ds_0^2=H(r,R)^{-1/2}(-dt^2)+H(r,R)^{1/2}\left[\sum_{i=7}^9d(\theta^iR)^2+dr^2+r^2d\Omega_5\right] \nonumber \\
&&e^\sigma=g_{0} H(r,R)^{3/2}
\eea
where the harmonic function is given by,
\be
H(r,R)=1+g_{0}l_s^7\sum_{n=1}^N\sum_{n_i}\left(\frac{Q_n}{[(x_7-x_{n_7}+n_7R)^2+...+\mid r-r_n \mid^2]^{7/2}}\right) \nonumber
\ee
where $g_0$ is the string coupling constant, $l_s$ is the string length, $Q_n$ is the charge of the D0-brane, $R_i$ are the radius of the torus and $r$ is the radius on spherical coordinates for the $M^6$ space time.

At distances much longer that $R_i$, we can Poisson resum the expression for the harmonic function $H$, obtaining,
\be
H(r,R_i)=1+{g_0l_s^7k_3 \over r^4\Pi_iR_i}
\ee
where $k_3$ stand for the irrelevant constant factors not display on the forthcoming discussions. On this solution we take the NH limit, by defining a triple ``blow up'', 
\begin{eqnarray}
&&\alpha'\rightarrow 0\;\;\;,\;\;\;U=\frac{r}{\alpha'} \nonumber \\
&&u_i=\frac{R_i}{\alpha'}\;\;\;,\;\;\;g^2_{YM}=g_0\alpha'^{-3/2}
\end{eqnarray}
where $g_{YM}$ is defined as usual up to a $2\pi$ factor, irrelevant for our forthcoming discussions.

Note that the new variables $v_i,g_{YM}$ correspond properly speaking to a blow up of the point $R_i=0$ and $g_0=\infty$ on the moduli space of the above metric. The metric and dilaton for the resulting NH geometry is,
\begin{eqnarray}
&&ds_0^2= \alpha'\left[\frac{U^2}{R(N,u_i)^2}(-dt^2)+\frac{R(N,u_i)^2}{U^2}\sum d(\theta_iu_i)^2+\frac{R(N,u_i)^2}{U^2}dU^2+ \right. \nonumber \\
&&\;\;\;\;\;\; \left. \;\;\; + R(N,u_i)^2d\Omega_5\right] \nonumber \\
&&e^\sigma=\left(\frac{g_{0}\alpha'^{-3/2}}{\prod u_i}\right)R(N,u_i)^3\frac{\prod u_i}{U^3}
\end{eqnarray}
where now we have defined $R(N,u_i)=\left[\left(\frac{g_{0}\alpha'^{-3/2}k_3N}{\prod_3 u_i}\right)\right]^{1/4}$.

Let us consider the T-dual theory of the above geometry, by using the Busher rules for the metric and dilaton in the presence of Ramond fields. In future sections we will examine this procedure more closely, but for the actual proposes it is good enough to follow the usual rules \cite{bus1}. Once this is done we get the metric,
\begin{eqnarray}
&&ds_3^2= \alpha'\left[\frac{U^2}{R(N)^2}(-dt^2+\sum d(\theta_i\frac{1}{u_i})^2 )+\frac{R(N)^2}{U^2}dU^2+R(N)^2d\Omega_5\right] \nonumber \\
&&e^\phi=g_3 
\end{eqnarray}
where $R(N)=\left(g_{3}Nk_3\right)^{1/4}$. This metric corresponds to a D3-brane NH geometry on a Torus $T^3$. Therefore, we have recovered the NH geometry of the D3-brane. Again, note that the D3-brane is wrapped around the $T^3$. In the limit of small $u_i$ we recover the flat D3-brane. This limit is related to a small expectation value for the Higgs fields on the D0-brane side, rather than a small compact circle!.

The above type of computation can be reproduced for any D-brane. Therefore we can obtain the NH T-dual geometry corresponding to any T-dual D-brane. Hence we can interrelate the fields and variables of the ``other'' NH geometries with the well understood D3-brane geometry, characterising the global structure of the T-dual spaces. This can be summarised on the following diagram,      

\begin{picture}(400,410)(-35,50)
\put (80,400){D3}
\put (80,300){D2}
\put (80,200){D1}
\put (80,100){D0}
\put (220,400){$AdS_5\otimes S^5$}
\put (220,300){$AdS_4\otimes S^5\otimes S^1$}
\put (220,200){$AdS_3\otimes S^5\otimes S^2$}
\put (220,100){$AdS_2\otimes S^5\otimes S^3$}
	\put (140,380){NH}
	\put (140,280){NH}
	\put (140,180){NH}
	\put (140,80){NH}
\put (260,350){T($S^1$)}
\put (260,250){T($S^1$)}
\put (260,150){T($S^1$)}
	\put (50,350){T($S^1$)}
	\put (50,250){T($S^1$)}
	\put (50,150){T($S^1$)}
\put (110,405){\vector(1,0){100}}
\put (110,305){\vector(1,0){100}}
\put (110,205){\vector(1,0){100}}
\put (110,105){\vector(1,0){100}}
	\put (85,390){\line(0,-1){70}}
	\put (85,290){\line(0,-1){70}}
	\put (85,190){\line(0,-1){70}}
\put (250,390){\line(0,-1){70}}
\put (250,290){\line(0,-1){70}}
\put (250,190){\line(0,-1){70}}
	\put (40,405){\line(0,-1){200}}
	\put (40,405){\line(1,0){20}}
	\put (40,205){\line(1,0){20}}
	\put (5,300){T($T^2$)}
\put (0,405){\line(0,-1){300}}
\put (0,405){\line(1,0){20}}
\put (0,105){\line(1,0){20}}
\put (-40,250){T($T^3$)}
\end{picture}\\
The vertical lines mean T-duality and horizontal lines mean NH limits.

To proceed with the research program, we recall that the NH geometries under study have to satisfy some inequalities to show holography. Basically, we need to rely on perturbative quantum gravity calculations, hence we demand small values for the Ricci scalar $\Re$, and small values of the dynamical string coupling $e^{\sigma}$. For the D0-brane case these restrictions translate into the following inequalities,
\begin{eqnarray}
\Re \approx \frac{1}{R^2}\ll 1 \rightarrow \frac{g_o\alpha'^{-3/2}}{\prod u_i}N\gg 1\;\;\;  \;\;\;  \frac{g_{YM}}{\prod u_i} N \gg 1 \\
e^{\sigma} \approx \left(\frac{g_{0}\alpha'^{-3/2}}{\prod u_i}\right)\left[\left(\frac{g_{0}\alpha'^{-3/2}N}{\prod u_i}\right)\right]^{3/4}\frac{\prod u_i}{U^3} \ll 1
\end{eqnarray}
The solution of these inequalities must be treated with some care. If we are considering the NH geometries in a regime where there is a T-dual map to other NH geometry, we have to add another inequality, stating that the ```would be'' string coupling constant on the T-dual geometry is also small. In our case, this is precisely the string coupling constant of the D3-brane i.e. $g_3 \ll 1$. Once this is taken into account, the above inequalities imply that,
\begin{eqnarray}
N \;>>1\;\;\;, U\gg Ru_i 
\end{eqnarray}
Therefore, to have holography in this framework, we should demand that:
\begin{itemize}
\item $N$ is large  (trusting sugra)
\item $g_{YM}$ small  (pertubative string theory)
\item $u_i$ small   (big range on the $U$ variable on the holography)
\end{itemize}

If the above conditions on the NH geometry are realised, we can see what is really going on. { \it The metric for the D0-branes approaches a $AdS_2$ times the five-sphere times some perturbations. We know that we have holography on this configuration (as this metric is T-dual to a D3-brane metric), Therefore we are ready to say that we have holography on the bulk space, when the space time geometry for a general Dp-brane takes the form  $AdS_{5-s}\otimes S^5\otimes T^s$ and the torus is really a perturbation of the other terms of the metric. The resulting structure is very natural from a holographic point of view, we understand this holography as the ``bulk'' space is on these cases $AdS$ space time}. 

Note that for the D0-brane case we have an $S^1$ frontier and $AdS_2$ bulk space, in the maximal holographic case (when the radius of the torus is infinitesimal), but as long as $u_i$ gets bigger, the holography is understand only from a given value of $U$ called it $U_0$, to infinity. What happens on the area defined for $U<U_0$ is not well understand and therefore we left it out of the analysis.

The above diagrams show the area of validity for the Maldacena's conjecture when a T-duality map is considered to a D3-brane NH geometry.
\vspace{10mm}

\begin{picture}(400,400)(20,0)
\put (70,400){D0-geometry}

\put (100,330){\oval(120,80)}
\put (100,330){\oval(110,70)}
\put (100,330){\vector(1,1){35}}
\put (110,335){$U_0$}
\put (140,270){$AdS_2$}
\put (155,280){\vector(0,1){25}}

\put (100,200){\oval(120,80)}
\put (100,200){\oval(60,30)}
\put (100,200){\vector(1,1){15}}
\put (110,200){$U_0$}
\put (140,140){$AdS_2$}
\put (155,150){\vector(-1,1){25}}

\put (100,70){\oval(120,80)}
\put (120,20){$AdS_2$}

\put (270,400){D3-geometry}

\put (300,330){\oval(120,80)}
\put (320,280){$AdS_5$}

\put (300,200){\oval(120,80)}
\put (320,150){$AdS_5$}

\put (300,70){\oval(120,80)}
\put (320,20){$AdS_5$}

\put (10,20){\vector(0,1){370}}
\put (15,20){$U_0=0$}
\put (15,380){$U_0=\infty$}

\put (170,330){\line(1,0){65}}
\put (170,200){\line(1,0){65}}
\put (170,70){\line(1,0){65}}

\put (180,310){$T-dual$}
\put (180,180){$T-dual$}
\put (180,50){$T-dual$}

\end{picture}

Here the space between the ovals represents $AdS$ bulk space, The frontier is represented by the oval itself, In the case of $AdS_2$ we have an $S^1$ and in the case of $AdS_5$ we have $S^1\otimes T^3$.

%%%%%%%%%%%%%%%%%%%%%%%%%%%%%%%%%%%%%%%%%%%%%%%%%%%%%%%%%%%%%%%%%%%%%%%%%%%%%%%

\section{T-duality and Gauge theories}

%%%%%%%%%%%%%%%%%%%%%%%%%%%%%%%%%%%%%%%%%%%%%%%%%%%%%%%%%%%%%%%%%%%%%%%%%%%%%%%

To begin with, let us start considering a D3-brane living in a ten dimensional space time with one of the orthogonal coordinates compactified on a circle $S^1$ of radius $R$. The corresponding metric is given by
\bea
ds_3^2&=& H(r,R)^{-1/2}(dx_{||}^2)+H(r,R)^{1/2}(d(\theta R)^2+dr^2+r^2d\Omega_4) \nonumber \\
e^\phi&=&g,
\label{d3b}
\eea
with $x_{||}$ standing for world volume coordinates and the harmonic function $H$ given by
\be
H(r,R)=1+gl_s^4\sum_{n=1}^N\sum_{n_i}\left(\frac{Q_n}{[(y_9-y_{n_9}+n_9R)^2+\mid r-r_n \mid^2]^{4/2}}\right),
\label{har1}
\ee
where $g$ is the string coupling constant, $l_s$ is the string length, $Q_n$ is the charge of the D3-brane, $R$ is the radius of the circle $S^1$ and $r$ is the radius on spherical coordinates for the $\Re^5$ space time. From now on we will ignore any constant and will work with meaningful variables in the discussion, and always at large N.

At distances much longer than $R$, we can Poisson resum the expression (\ref{har1}) obtaining
\be
H(r,R)=1+{l_s^4g \over Rr^3}.
\label{har2}
\ee
This is a good solution as far as $R$ is small enough.

If we are interested in the NH limit \cite{mal1} of this metric we would be forced to defined this limit by performing a double ``blow up'', namely
\bea
\left ( \alpha'\rightarrow 0\;\;\;,\;\;\;U\equiv\frac{r}{\alpha'}= constant\;\;\;,\;\;\;v\equiv\frac{R}{\alpha'}=constant\right ).
\label{nh1}
\eea
Notice that the new variable $v$ correspond properly speaking to a blow up of the point $R=0$ in the moduli of the target space time metric (\ref{d3b}) with the harmonic function of (\ref{har2}). After performing the blow up (\ref{nh1}) we get the metric
\bea
ds_3^2&=& \alpha'\left[ \frac{U^{3/2}v^{1/2}}{g^{1/2}}(dx_{||}^2) + \frac{g^{1/2}}{U^{3/2}v^{1/2}}dU^2+ \frac{g^{1/2}U^{1/2}}{v^{1/2}}d\Omega_4 +\frac{g^{1/2}v^{3/2}}{U^{3/2}}d\theta^2 \right]\nonumber \\
e^\phi&=&g.
\label{m2} 
\eea
Note that the dilaton field for this solution is constant. The topology of the space time (\ref{m2}) is that o a fibration of a circle $S^1$ and a sphere $S^4$ on the space defined on the coordinates $(x^\alpha,U)$ with the corresponding radius
\bea
R_{S^1}&=&({gv^3 \over U^3})^{1/4}, \nonumber \\ 
R_{S^4}&=&({ gU \over v})^{1/4},
\label{rad1}
\eea
which depends on the moduli $v$. It is easy to see that this space time admits only $16$ real supercharges. The simplest way to understand this is by observing that $1 \over 2$ of the Killing spinors for the near horizon geometry of the D3-brane are projected out once we compactify a transverse direction (we will come back to this point later on). 

From the QFT point of view we should expect the metric (\ref{m2}) to be related to a SYM theory in $3+1$ dimensions, with $16$ real supercharges and with a peculiar R-symmetry given by the isometries of $S^4 \otimes S^1$. The type of strings living on the space time (\ref{m2}) is Type IIB.

Consider next the candidate for a T-dual geometry, namely the D4-brane compactified on a $S^1$ of radius $R$, on its world volume. This solutions is given by
\bea
ds_4^2&=& H^{(-1/2)}(dx_{||}^2+d(\theta R)^2)+H^{(1/2)}(dr^2+r^2d\Omega_4) \nonumber \\
e^\phi&=&gH^{-1/4}, 
\eea
with $x_{||}$ expanding the non-compact four dimensional part of the world volume of the D4-brane, and the corresponding harmonic function
\be
H=1+\frac{gl_s^3}{r^3}.
\ee
Taking the NH limit, defined by again a double blow up
\bea
\left ( \alpha'\rightarrow 0\;\;\;,\;\;\;U\equiv\frac{r}{\alpha'}=constant\;\;\;,\;\;\;g_5^2\equiv g\alpha'^{1/2}=constant \right ),
\eea
we get the metric
\bea
ds_4^2&=& \alpha'\left[\frac{U^{3/2}}{g_5}(dx^2_{||}+(d\theta R)^2 )+\frac{g_5}{U^{3/2}}dU^2+g_5U^{1/2}d\Omega_4\right] \nonumber \\
e^\phi&=&g_5^{3/2}U^{3/4}.
\label{nh2}
\eea
Note that this time the dilaton is not constant. The number of supersymmetries in this brane is again $16$ real supercharges, however the QFT interpretation is a bit different. In this case we have a $D=5$ SYM theory living on $\Re^4\otimes S^1$. The corresponding radius of the  four-sphere and the circle are
\bea
R_{S^4}&=&(g_5^2U)^{1/4}, \nonumber \\ 
R_{S^1}&=&(\frac{U^{3}R^4}{g_5^2})^{1/4}.
\label{rad2}
\eea
\begin{figure}
\begin{center}
\includegraphics[angle=-90,scale=0.4]{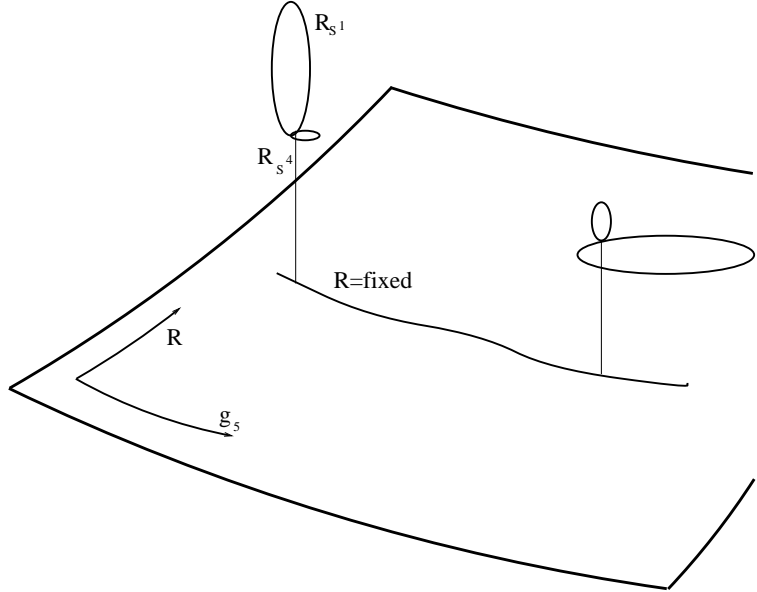}
\caption{Moduli space for near horizon D4-brane.}
\label{fig12}
\end{center}
\end{figure}
Before proceeding any further, let us clarify the picture we have (see fig. \ref{fig12}). In the D3-brane case we obtained the near horizon geometry as the result of a limit where one of the moduli $g$  was left constant, but we allowed $R$ to varied such that, at $R=0$, we created a divisor $v$. These two variables defined our moduli $(v,g)$. The resulting geometry is that of a base space expanded by the coordinates $(x,U)$ and fibers $S^4$ and $S^1$ with the radius of equation (\ref{rad1}). In the D4-brane, we obtained the near horizon geometry as the result a another limit where one of the moduli $R$ is maintained constant while the other $g$ varies such that at the infinite point we create a new divisor $g_5$. These two variables define the new moduli $(R,g_5)$. Again the geometry obtained is that of a base manifold expanded by the coordinates $(x,U)$ and fibers $S^4$ and $S^1$ with the radius of equation (\ref{rad2}). In order to identify both metrics (\ref{nh1},\ref{nh2}) by T-duality we must require the following relation between the different moduli,
\bea
g={g_5^2 \over R}, \nonumber \\
v={1 \over R}.
\label{td1}
\eea
Provided that this relation holds, we can perform the T-duality transformation following the Buscher rules \cite{bus1}. In principle we could run into difficulties if too many singular point appear on the fibration, so that T-duality could loose its natural meaning. This T-dual map is well defined all over the base manifold. Actually we only have singularities at $U=0$ and $U=\infty$, but both points are related to wrong coordinate patches rather than real singularities. Therefore this T-dual map is a very trivial example of fiber-wise T-duality \cite{asp1}. The map described above is defined by T-duality, and effectively acts between the two moduli.

By now we have a neat relation between the bare coupling constants of both gauge theories in the two near horizon metrics. On the other hand, the dilaton behavior is usually associated with the value of the corresponding running coupling constant i.e. the effective gauge coupling constant. Therefore, to obtain the effective coupling of the compactified gauge theory on the world volume of the D4-brane, we consider the ratio of the effective coupling constant of the five dimensional gauge theory $g_5$ squared, with the effective radius of compactification namely the radius of the $S^1$ given on equation (\ref{rad2}), hence we get
\be
g_{4_{eff}}^2\equiv {g_{5_{eff}}^2 \over R_{eff}}.
\label{geff}
\ee
Then after solving for the moduli variables $(v,g)$ we obtain
\be
 g_{4_{eff}}^2=g.
\label{gf}
\ee
Therefore, the effective coupling constant of the gauge theory we are studying from the point of view of the near horizon D4-brane has the same running behaviour as the the gauge theory on the near horizon D3-brane, as should be expected invoking duality. Note that equation (\ref{geff}) was obtained by plausible physical relations, however this equation is nothing more than the rule for changing the dilaton under T-duality!

On the other hand, the super Yang Mills theory on the D4 brane is not renormalizable, we can trust it only at low energies. This aspect of the gauge theory can be seen from the gravitational point of view. Note that, for the geometry (\ref{nh2}), the dilaton grows for large $U$. Actually, we can trust this solution as long as $U \geq 1/g_5^2$, after this point we should think in terms of M-theory.

The other possibility we have is to consider the D3-brane wrapped on a circle $S^1$ of radius $R$. This time the NH geometry is defined by the limit
\be
\left ( \alpha'\rightarrow 0\;\;\;,\;\;\;U\equiv\frac{r}{\alpha'}= constant \right ),
\ee
where ($g,R$) are kept constant on the process. Note that this time we don't have the double blow up of the above cases. The resulting metric is given by
\bea
ds_{3}^2&=& \alpha'\left[ \frac{U^2}{g^{1/2}}( d(R\theta)^2+dx_{||}^2)+\frac{g^{1/2}}{U^2}dU^2 +g^{1/2}d\Omega_{5})\right] \nonumber \\
e^\phi&=&g.
\label{nh3}
\eea
Again the dilaton is constant, and the topology is that of a fibration of a circle $S^1$ and the sphere $S^5$, on the space defined by the coordinates ($x^{\alpha},U$), with the corresponding radius
\bea
R_{S^1}&=&{UR \over g^{1/4}}, \nonumber \\ 
R_{S^5}&=&g^{1/4}.
\label{rad3}
\eea
This configuration also only admits 16 real supercharges. The situation on the field theory should be that of a $D=4$ SYM theory living on $\Re^3\otimes S^1$ with 16 real supercharges and R-symmetry contained on $SO(6)$.

The T-dual NH background is the result of first, a Poisson resume of the D2-brane solution with a transverse direction compactified on a small circle $S^1$ of radius $R$, second its NH limit defined by the triple blow up, here showed
\bea
 \alpha'\rightarrow 0\;\;\;&,&\;\;\;U\equiv\frac{r}{\alpha'}=constant, \nonumber \\
v\equiv { R \over \alpha'}\;\;\;&,&\;\;\;g_3^2\equiv g\alpha'^{-1/2}=constant. 
\eea 
The resulting metric and dilaton are given by
\bea
ds_{2}^2&=& \alpha'\left[ \frac{U^2v^{1/2}}{g_3}(dx_{||}^2)+\frac{g_3}{U^2v^{1/2}}dU^2+{g_3 \over v^{1/2}}d\Omega_{5} + \frac{g_3v^{3/2}}{U^2}d\theta^2 \right] \nonumber \\
e^\phi&=& {g_3^{5/2} \over Uv^{1/4}}.
\label{nh4}
\eea
This time we have space time with 16 real supercharges, the metric defines a fibration on a circle $S^1$ and the sphere $S^5$ on the base space expanded by the coordinates ($x^\alpha,U$) and the field theory point of view should be that of a SYM theory on $2+1$ dimensions, with 16 real supercharges, and R-symmetry contained in $S0(6)\otimes U(1)$. The corresponding radius of the fibers are 
\bea
R_{S^1}&=&{g_3^{1/2}v^{3/4} \over U^{1/2}},\nonumber \\ 
R_{S^5}&=&{g_3^{1/2} \over v^{1/4}}.
\label{rad4}
\eea
To perform the T-duality map between this two metrics, we require to identify the moduli as follows,
\bea
g={g_3^2 \over v}, \nonumber \\
R={1 \over v}.
\label{td2}
\eea
Similar remarks about the validity of T-duality of the D3-brane and the D4-brane are applicable to this case.

In general, we can consider the above type of compactification of the D3-brane on a $T^{3-s}$, which takes us to the near horizon D(s)-brane on a $T^{3-s}$ in the perpendicular coordinates. The metric of these geometries looks like $AdS_{s+2}\otimes S^5\otimes T^{3-s}$. When the holography map between the gauge theories and the bulk geometries is defined \footnote{Recall that the holographic conjecture of Maldacena only applies within a range of validity of $U$ for the case in which p is different from 3} the radius of the torus is very small. Also we find that when $R\rightarrow \infty$, we recover on these D(s)-branes the full range of running for $U$ ($0,\infty$), while for small $R$ we are forced to stay at big values of the holographic variable $U$. 

It is well known that T-duality breaks supersymmetry in some cases \cite{alv1} and our previous metrics are not exceptions to this phenomenon. Note that we are relating theories with 16 real supercharges, but we already showed that the near horizon D3-brane shows enhancement of supersymmetry. The matching condition for the number of supersymmetries is given by the fact that the compact direction on the D3-brane (on both cases) eliminates the possibility of that enhancement, while the other Dp-branes don't show the enhancement at all. Let us study a bit more carefully this point.

 In general the relevant system of Killing spinor equations for the Dp-brane ansatz is given by
\bea
\delta \lambda &=& H^{-1/4}\Gamma^r\partial_r\phi\epsilon+{(3-p)e^\phi(\partial_r H)\Gamma^r \over 4H^{8-p\over 4}}
\Gamma_0 ... \Gamma_p \epsilon^{'}  \nonumber \\
\delta \psi_{\alpha} &=& \partial_{\alpha} \epsilon + {(\partial_r H)\over 8H^{3\over 2}}\Gamma^r \Gamma_{\alpha}\epsilon + {e^\phi(\partial_r H)\over 8H^{9-p\over 4}}\Gamma^r \Gamma_{\alpha}\Gamma_0 ... \Gamma_p \epsilon^{'} \nonumber \\
\delta \psi_r &=& \partial_r \epsilon -{e^\phi(\partial_r H) \over 8H^{7-p \over 4} } \Gamma_0 ...\Gamma_p \epsilon ^ {'} \nonumber
\label{susy1}
\eea
\begin{equation}
\epsilon_{(0,4.8)} ^{'} = \epsilon 
\qquad
\epsilon_{(2,6)} ^{'} = \Gamma_{11} \epsilon
\qquad
\epsilon_{(-1,3,7)} ^{'} = \imath \epsilon
\qquad
\epsilon_{(1,5)} ^{'} = \imath \epsilon^{\ast}
\end{equation}
Where we have solved for the Dp-brane ansatz using the split $M=(\alpha,r,\theta)$ where  $(r,\theta)$ are perpendicular coordinates to the brane, also $\epsilon $ is a 32-component spinor, and $\omega$ is the spin connection \cite{ber1}. Note that we have on propose leave the dilaton unspecified in terms of $H$. It is important to recall that the dilatino equation is crucial as projects out one half of the supersymmetries. To define the NH limit of the above system, it is convenient to give a more geometrical meaning to the dilatino, as a part of the component of the relevant gravitino in an embedding supergravity like D=11 supergravity for the type IIA D-branes solutions or in a F-theory framework for type IIB D-branes. Once this is done (for example in the 11D case) the NH limit of the above system gives,
\bea
\delta \lambda &=& \alpha'^{1/6}\left[{(3-p)(\partial_u h)\Gamma^u \over 4h^{5\over 4}}[\epsilon + \Gamma_0 ... \Gamma_p \epsilon^{'}]\right] \nonumber \\
\delta \psi_{\alpha} &=& \partial_{\alpha} \epsilon + {(\partial_u h)\over 8h^{3\over 2}}\Gamma^u \Gamma_{\alpha} [\epsilon + \Gamma_0 ... \Gamma_p \epsilon^{'}] \nonumber \\
\delta \psi_u &=& \partial_u \epsilon -\left({\partial_u h \over 8h}\right)
\Gamma_0 ...\Gamma_p \epsilon ^ {'}  \nonumber
\eea
where $h=g^2_{YM}/U^{7-p}$.

In this case the dilatino equation goes to zero in the near horizon limit, giving no constraints. Nevertheless the consistency conditions corresponding to the gravitino equation, contains the dilatino equation among others. After the usual decomposition of the killing spinors, defined by the canonical projector on this ansatz, we find that one half of the supersymmetry is always preserved if the dilatino constraint is satisfied , but the other half is conserved only if the function $h$, behaves as
\be
h \propto {1 \over U^4},
\ee
therefore we found the possibility of enhancement for $p=3$ only\footnote{In principle we could try to generalized the function $h$ to get this wanted behavior, in fact there is an approach debt to Jevisky and Yoneya \cite{jevi1}, where the string coupling constant $g$ is allowed to depend on the coordinates of the bulk geometry, such that the function $h$ behaves under a conformal symmetry as before. Nevertheless although the right transformation  under conformal symmetry is archived by this type of generalizations, only if $g$ depends on the holographic variable $U$ we will recover the decider behavior for $H$. In any case, a new type of symmetry called the ``Generalized Conformal Symmetry'' (GCS) is defined. We found that once the string coupling constant $g$ is not a ``constant'', the connection between the geometry and a non-linear sigma model is lost, as we don't have a solution for the zero value of the sigma model beta function any more.\\   

The situation for the near horizon geometries we are considering is a bit different, in fact the solution for a D3-brane on a torus $T^s$ on its world volume and its treatment to the NH limits gives a function $h=\frac{g_s^2}{\Pi_sR^iU^4}$ with the right dependence on $U$. The full GCS of the corresponding near horizon geometry is recovered, as the action of the conformal transformations of the near horizon D3-brane induce by T-duality on the NH Dp-brane. For example we have, for the infinitesimal transformation corresponding to a dilation acting on $AdS_5$
\bea
\delta_\rho U&=& 2\rho U \nonumber \\
\delta_\rho x^a &=& -2\rho x^a \nonumber \\
\delta_\rho g &=& 0
\eea
for the will-be coordinates on the T-dual geometries this transformation translate into,
\be
\delta_\rho (R_i)=  -2\rho(R_i)
\ee
Also the action on the string coupling constant gives information,
\be
\delta_\rho g = 0 \rightarrow \delta_\rho\left( g_{YM}^2\prod_sR^ii\right)=0  \rightarrow \delta_\rho g_{YM}^2 = 2s\rho g_{YM}^2 
\ee
Therefore we get the following transformations,
\bea
U & \rightarrow & \lambda U \nonumber \\
x^a & \rightarrow & \lambda^{-1}x^a \nonumber \\
g_{YM}^2 & \rightarrow & \lambda^s g_{YM}^2
\eea
Note nevertheless, that this induce transformation is of a very queer type as it acts on the moduli space and on the space time variables at the same time. Under this circumstances we don't believed this should be called a symmetry, at least in the usual sense. Also on the T-dual D-brane, related to the D3-brane, we have no enhancement of supersymmetry due to the holonomy involved in this solutions. This is also understand as an efect of T-duality breaking the will-be new supersymmetries.}.

It is important to notice that the supersymmetries which are broken by T-duality correspond to those which get enhanced in the particular case of the D3-brane, i.e. the bilinear associated with the broken Killing spinors, are the conformal Killing vectors.

In our previous analysis we have considered the two dual pairs (D3/D4) and (D2/D3). Both start with the D3-brane, with the only difference that the compactified dimension is or is not in the transversal direction. Let us study a bit more carefully both pairs. In the (D3/D4) case, the D4-metric is characterized by two different moduli $(g_5,R)$. After the work in Matrix theory \cite{sei1}, it is natural to interpret $g_5$ as related to the eleventh dimension of M-theory and therefore to consider this metrics as coming from a compactification of M-theory on a $T^2$, with sizes determined by the two moduli $g_5$ and $R$. Also it is well known \cite{sus2} that in the limit where the volume of the two torus goes to zero, we should recover type IIB theory. This mechanism implies the dynamical generation of a ``quantum'' dimension with the corresponding Kaluza Klein modes associated with the membrane wrapped on the two torus. When we apply this mechanism to our case it is natural to expect to get in the limit of zero volume for the two torus ($Vol(T^2)\rightarrow 0$) the type IIB D3-brane metric of equation (\ref{m2}). Using the relations (\ref{rad1}) and (\ref{td1}) we observe that the limit $Vol(T^2)\rightarrow 0$ corresponds to a ten dimensional type IIB theory, where the extra ``quantum'' dimension is the $S^1$ circle in the limit $v\rightarrow\infty$, with the radius given by (\ref{rad1}). The up-lift of the D4-brane to M-theory gives us on M5-brane wrapped on a circle determined by the value of $g_5$. In addition to this, in our case we wrap the M5-brane on another circle defining the two torus characterized by the two moduli $(g_5,R)$. In the limit $Vol(T^2)\rightarrow 0$ what we get is the M5-brane wrapped on a two torus of zero volume, that produce a D3-brane with the extra dimension defining the transversional circle in the metric (\ref{m2}).

Hence, the theory on the D4-brane gets embedded in the six dimensional ($2,0$) theory on the M5-brane. As is well known for the M5-brane, we get enhancement of supersymmetry and therefore we can say that the D4-brane theory will flow to a conformal point in strong coupling. {\it In other words what we observe is that, once we break the superconformal generators by T-duality, the resulting theory naturally flows to recover the supersymmetry by up lifting to M-theory}. 

Let us now consider the T-dual pair (D2/D3). This is very similar to the previous case. In the D2-brane metric the moduli is characterized by $(g_3,v)$, which again should be interpreted as M-theory compactified on a two torus of size $g_3$ and $v$. In the limit when the volume of the two torus goes to zero, we should recover the type IIB-picture by exactly the same mechanism described above. The D2-brane is now up lifted to a M2-brane but contrary to what happens in the (D3/D4) case, the extra ``quantum'' dimension becomes now part of the world volume dimension of the T-dual D3-brane. More precisely what we observed is that the compact ``world volume'' dimension in the metric (\ref{nh3}) is the extra ``quantum'' dimension in the type IIB that we get when we compactify M-theory on the two torus characterized by ($g_3,v$). {\it In other words, what  we observe is that the T-dual description in  (\ref{nh3}) of the uplifted D2-brane is a three dimensional theory becoming four dimensional for ``strong coupling'' $v\rightarrow 0$ in equation (\ref{td2})}.

As before with the (D3/D4) pair we also observe here that the theories flow to reach superconformal invariance. Notice that these two dual pairs saturate the known examples of superconformal theories namely the D3-brane, M2-brane and M5-brane. The (D3/D4) pair is related to the M5-brane and the (D2/D3) pair to the M2-brane.
\begin{figure}
\begin{center}
\includegraphics[angle=-90,scale=0.4]{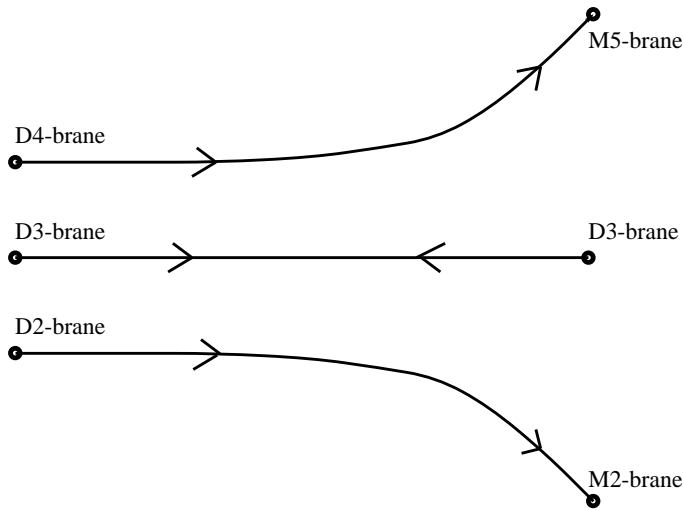}
\caption{Flow to the superconformal QFT.}
\label{fig13}
\end{center}
\end{figure}

The previous conjecture is in contrast to the mechanism suggested in \cite{wit1} for solving the cosmological constant problem. In that we can start with a three-dimensional theory that is expected to flow in strong coupling to a four-dimensional theory. Massive particles in three dimensions are associated with conical geometries, when some amount of supersymmetry is broken. The suggested solution to the cosmological constant problem, is based on the assumption that these supersymmetries are not restored in the strong coupling four-dimensional limit. In our case, we  have simply studied supersymmetry generators associated with conformal transformations that are the ones naturally broken by the action of T-duality, and we find they are restored in the up lifted ``M-theory'' limit.  

The general picture emerging from the previous discussion is that once we start with a superconformal theory, T-duality generally breaks the supersymmetries associated with the superconformal transformations. However the T-dual theory tends to flow to recover these supersymmetries broken by T-duality  up lifting to M-theory.

To be more precise, starting with a D3-brane with the world volume compactified on a circle of radius $R$, we break for finite radius the supersymmetries associated with those Killing spinors depending on world volume coordinates. Those are associated with the enhanced  supersymmetry. In order to decide if T-duality breaks supersymmetry or not, we perform a T-duality to a D2-brane. Once we have done that, we send the radius $R$ to infinity. In this limit we recover for the D3-brane the whole superconformal algebra. Then if T-duality is not breaking supersymmetry we should find that the T-dual of  the $R\rightarrow\infty$ limit possesses enhanced superconformal invariance. In fact this is what happened. By relation (\ref{td2}), when $R\rightarrow\infty$ then $v\rightarrow 0$ and $g_3^2\rightarrow 0$ for finite $g$, but $g_3^2$ can be interpreted as $1/\Delta$ for $\Delta$ the size of the eleventh dimension. Thus the D2-brane becomes uplifted to M2-brane recovering the superconformal transformations. In a certain sense M-theory is there to work out the breaking of supersymmetry induced by T-duality.

\vspace{12pt}
{\bf Acknowledgements}
\vspace{12pt}

The author would like to thanks the C. Gomez and E. Alvarez for their valuable comments on various subjects treated on this work, Also this work was partially supported by the Venezuelan Government.

\end{document}